\newcommand{\bea}{\begin{eqnarray}}
\newcommand{\eea}{\end{eqnarray}}
\newcommand{\nn}{\nonumber}

\documentclass{ws-ijmpd}
\usepackage[super,compress]{cite}
\begin{document}

\markboth{A. Kuiroukidis}{
A class of simple bouncing and late-time accelerating
cosmologies in $f(R)$ gravity
}
%{Instructions for Typing Manuscripts (Paper's
%Title)}

\catchline{}{}{}{}{}
%
%%%%%%%%%%%%%%%%%%%%%%%%%%%%%%%%%%%%%%%%%%%%%%%%%%%%%%%%%%%%%%%%%%%%
\title{A class of simple bouncing and late-time accelerating
cosmologies in $f(R)$ gravity
%\footnote{
%For the title, try not to use more than
%3 lines. Typeset the title in 10 pt roman, uppercase and
%boldface.}
}

\author{A. Kuiroukidis
%\footnote{
%Typeset names in 8 pt roman, uppercase. Use the footnote to indicate the
%present or permanent address of the author.}
}

\address{Department of Informatics,\\
Technological Education Institute of Central Macedonia,\\
GR 621 24, Serres Greece\\
%City, State ZIP/Zone, Country
%\footnote{
%State completely without abbreviations, the affiliation and
%mailing address, including country. Typeset in 8 pt italic.}\\
apostk@teicm.gr}

%\author{Second Author}

%\title{INSTRUCTIONS FOR TYPESETTING
%MANUSCRIPTS\footnote{For the title, try not to use more than 3 lines.
%Typeset the title in 10~pt Times roman, uppercase and boldface.}  }

%\author{FIRST AUTHOR\footnote{Typeset names in
%8~pt roman, uppercase. Use the footnote to indicate the
%present or permanent address of the author.}}

%\address{University Department, University Name, Address\\
%City, State ZIP/Zone,
%Country\footnote{State completely without abbreviations, the
%affiliation and mailing address, including country. Typeset in
%8~pt Times italic.}\\
%first\_author@university.edu}

%\author{SECOND AUTHOR}

%\address{Group, Laboratory, Address\\
%City, State ZIP/Zone, Country\\
%second\_author@group.com}

\maketitle

\begin{history}
\received{Day Month Year}
\revised{Day Month Year}
\end{history}

\begin{abstract}
We consider the field equations for a flat FRW cosmological model,
in a generic $f(R)$ gravity model and cast them into a,
completely normalized-dimensionless, system of
O.D.Es for the scale factor and the function $f(R)$, with respect to
the scalar curvature $R$. It is shown that under reasonable
assumptions one can produce simple analytical and numerical solutions
describing bouncing cosmological models where in addition there are
late-time accelerating. Possibility of extending these results
is briefly discussed.
%The abstract should summarize the context, content
%and conclusions of the paper in less than 200 words. It should
%not contain any references or displayed equations. Typeset the
%abstract in 8 pt Times roman with baselineskip of 10~pt, making
%an indentation of 1.5 pica on the left and right margins.
\end{abstract}

\ccode{PACS numbers: 11.25.Pm}

%\tableofcontents

\section{Introduction}

Extended theories of gravity, alternate to Einstein's
General Relativity (GR), were proposed to solve the problem
of the standard cosmological model of GR and also in
relation to the quantization of gravity
(see e.g. [\citen{capo1}]-[\citen{fara1}]). Today the
standard cosmological model (SCM) also called the Big-Bang
cosmology is the most acceptable model of the Universe up
to this day. Among its major successes is the Hubble law
for the expansion of the Universe, the black body nature
of the cosmic microwave background (CMB) radiation and
the light element abundances predicted by it
(see e.g. [\citen{shab}] and references therein). However the
SCM includes a number of deficiencies such as the
problems of magnetic monopoles, gravitons and baryon asymmetry.
Also despite its success in describing the observable Universe
up to a hundredth of a second after its initial moment,
there exist unanswered questions regarding the initial state
of it, namely the horizon and flatness problems [\citen{kinn}].

Though the inflation concept has been introduced to solve
the above problems of the SCM, it suffers also from another
major unresolved problem, namely the existence of the initial
cosmological singularity, predicted by standard
singularity theorems [\citen{hawk}]. A singular initial
state is usually an extreme situation where physical observable
quantities such as temperature, mass-energy density and
curvature attain infinite values, a case which signifies
the breakdown of the underlying physical theory. Attempts to
remedy this situation include, among others, consideration
of cosmological models where in fact at the initial moment
(usually taken to be $t=0$) there exist finite values for the
physical quantities and in particular for the scale factor $a(t)$
of the evolution of the Universe and the scalar curvature $R$.
These are the so-called bouncing cosmological models, where
presumably the Universe undergoes a previous contracting phase
in its evolution.

There exists a vast literature on bouncing cosmological models,
mainly in the frame of extended theories of gravity, where the
GR action is replaced by a generic function of the scalar
curvature $f(R)$. A very incomplete list includes the following:
Bouncing cosmological models in the frame of $f(R)$ gravity and
bi-gravity appeared in [\citen{bamb}]. Bouncing cosmological
models from higher powers of gravity appeared in [\citen{clif}].
Static Einstein models in power-law form of the action, namely
in $f(R)$ gravity appeared in [\citen{gohe}]. Bouncing cosmological
models in Palatini $f(R)$ gravity appeared in [\citen{barr1}],
and also in [\citen{koiv}]. Bouncing cosmological models in
$F(R,\phi)$ gravity with phantom equation of state (EoS) appeared
in [\citen{faraj1}]. Isotropic and anisotropic bouncing
cosmological models in Palatini $f(R)$ gravity appeared in
[\citen{barr2}]. Bouncing cosmology from Lagrange-multiplier
modified gravity appeared in [\citen{fuca}], while an interesting
review of the Palatini approach to modified gravity such as
the case of $f(R)$ gravity appeared in [\citen{olmo1}]. Bouncing
cosmological models in the $f(R)$-Palatini variational
formalism appeared in [\citen{olmo2}], while $f(R,T)-$gravity and
its various phenomena and cosmological models appeared in
[\citen{hark1}]. Bouncing cosmological models in the frame of
Horava-Lifshitz $f(R)$ gravity appeared in [\citen{chak}], while
bouncing and late-time accelerating cosmological models,
in the Born-Infeld-like $f(R)$ gravity appeared in [\citen{fubr}].

According to recent cosmological observations in
terms of Supernovae Ia, large scale structure with the baryon
acoustic oscillations, cosmic microwave background radiation
and weak lensing, the current expansion of
the universe is accelerating (see e.g. the introduction of
[\citen{bamb}] and references therein). In this context, cosmological
models that attempt to incorporate both the early stage bouncing
behavior with late time accelerating phase were vastly considered
in the literature. An incomplete list includes e.g.
[\citen{dimi}]-[\citen{noji3}] and references therein.
In this paper we consider the field equations
for an $f(R)$ cosmology with metric described by Eq. (\ref{metric}),
as they appear in e.g. [\citen{capo1}]. We use an approach that,
to the author's knowledge, has not been explicitly followed in the relevant
literature. This consists of the assumption that we have a bouncing
cosmological model where the scalar curvature increases monotonically,
as we approach the initial moment $t=0$, in the contracting phase of
the Universe. This solution is then matched to its mirror-symmetric,
with respect to the time $t$, for the expanding phase of the Universe.
By a direct calculation the field equations are transformed
to a system consisting of a first order equation for the metric and a second
order equation for the function $f(R)$, with respect to the scalar curvature
$R$. These are then solved analytically, for some simple ansatze and numerically
also, resulting in bouncing cosmological models with late-time acceleration.

This paper is organized as follows: In section 2 we provide the
general setting, for an $f(R)$ cosmology, along with the field equations
describing it. In section 3 we explicitly give our ansatz and assumptions.
In section 4 we reduce the field equations to a completely normalized-dimensionless
system of O.D.Es for the scale factor of the metric and the function
$f(R)$ and we provide the class of simple analytical models that are
proposed. In section 5 we give an example of a cosmological model from
a numerical treatment of the above system. In section 6 we explain and
justify our construction regarding its smoothness, as a solution to the
field equations. Finally in section 7 we give a brief discussion of
the results.

%We use a flat FRW cosmological model with metric given by
%\bea
%\label{metric}
%ds^{2}=-dt^{2}+a^{2}(t)[dx^{2}+dy^{2}+dz^{2}]
%\eea
%where $a(t)$ is as usual the scale factor and $H=\dot{a}/a$ is the
%Hubble parameter.

%Contributions to {\it International Journal of Modern Physics D}
%are to be in American English. Authors are encouraged to have their
%contribution checked for grammar. American spelling should be
%used. Abbreviations are allowed but should be spelt out in full when
%first used. Integers ten and below are to be spelt out.  Italicize
%foreign language phrases (e.g.~Latin, French).  Upon acceptance,
%authors are required to submit their data source file including
%postscript files for figures.

%The text is to be typeset in 10 pt roman, single spaced with
%baselineskip of 13~pt. Text area (including copyright block)
%is 8 inches high and 5 inches wide for the first page.  Text area
%(excluding running title) is 7.7 inches high and 5 inches wide for
%subsequent pages.  Final pagination and insertion of running titles
%will be done by the publisher.

%\newpage

\section{The Setting}

We consider the $f(R)-$gravity action, with matter content [\citen{shab}]
\bea
\label{action}
S=S_{g}+S^{(m)}=\int d^{4}x\sqrt{-g}
\left[\frac{1}{2\kappa^{2}}f(R)+L^{(m)}\right]
\eea
where $\kappa^{2}=8\pi G$ and we use geometrical units by
setting $c=G=1$. The stress-energy tensor is given by [\citen{shab}]
\bea
\label{setens}
T_{\mu\nu}:=\frac{-2}{\sqrt{-g}}
\frac{\delta[\sqrt{-g}L^{(m)}]}{\delta g^{\mu\nu}}
\eea
Varying with respect to $g^{\mu\nu}$ we obtain the field
equations [\citen{capo1}]
\bea
\label{fieldeq}
f^{'}(R)R_{\mu\nu}-\frac{f(R)}{2}g_{\mu\nu}=\nabla_{\mu}\nabla_{\nu}f^{'}(R)
-g_{\mu\nu}\Box f^{'}(R)+\kappa^{2}T_{\mu\nu}
\eea
where a prime denotes derivative with respect to the scalar curvature $R$
and we use the notation $\Box :=g^{\mu\nu}\nabla_{\mu}\nabla_{\nu}$. The
trace of Eq. (\ref{fieldeq}) gives [\citen{capo1}]
\bea
\label{trace}
3\Box f^{'}(R)+R f^{'}(R)-2f(R)=\kappa^{2}T
\eea
where $T:=T^{\alpha}_{\alpha}:=g^{\mu\nu}T_{\mu\nu}$ is the trace
of the stress-energy tensor. Thus Eq. (\ref{trace}) shows that $f^{'}(R)$
is a dynamical-scalar degree of freedom of the theory.

We consider now a flat FRW metric
\bea
\label{metric}
ds^{2}=-dt^{2}+a^{2}(t)[dx^{2}+dy^{2}+dz^{2}]
\eea
where $a(t)$ is the scale factor and $H:=\dot{a}/a$ is the
Hubble parameter. The stress-energy conservation
$\nabla_{\nu}T^{\mu\nu}=0$ is automatically satisfied by the
field equations (\ref{fieldeq}). The field equations become [\citen{capo1}]
\bea
\label{fieldeq1}
3H^{2}&=&\frac{\kappa^{2}}{f^{'}(R)}
\left[\rho^{(m)}+\frac{Rf^{'}(R)-f(R)}{2}-3H\dot{R}f^{''}(R)\right]\nn \\
2\dot{H}+3H^{2}&=&\frac{-\kappa^{2}}{f^{'}(R)}
\left[P^{(m)}+f^{'''}(R)(\dot{R})^{2}+2H\dot{R}f^{''}(R)+\right.\nn \\
&+&\left.\ddot{R}f^{''}(R)+\frac{f(R)-Rf^{'}(R)}{2}\right]
\eea
namely the generalized Friedman and Raychaudhuri equations.

We use now a single perfect fluid with barotropic equation of state (EoS)
of the form $P^{(m)}=w\rho^{(m)}$. Thus, with the sign conventions
we adopt for the metric, we have
\bea
\label{setens1}
T^{\mu}_{\nu}=diag
\left(-\rho^{(m)},P^{(m)},P^{(m)}, P^{(m)}\right)
\eea
where $T=-\rho^{(m)}+3P^{(m)}=(3w-1)\rho^{(m)}$, while we
also have
\bea
\label{scalar}
R=6\left(\frac{\ddot{a}}{a}+\frac{\dot{a}^{2}}{a^{2}}\right)=6(\dot{H}+2H^{2})
\eea
for the scalar curvature of the cosmological spacetime metric
of Eq. (\ref{metric}).  For the rest of this paper we will be
considering pressureless dust, i.e. $w=0$ and from the stress-energy
tensor conservation $\nabla_{\nu}T_{\mu}^{\nu}=0$, we obtain
\bea
\label{dust}
\rho^{(m)}=\rho_{0}
\left(\frac{a_{0}}{a}\right)^{3},\; \; \; T=-\rho^{(m)}
\eea

\section{The model}

We now write the field equations (\ref{fieldeq1}) as [\citen{shab}]
\bea
\label{fieldeq2}
3H^{2}&=&\kappa^{2}\rho_{(eff)}\\
2\dot{H}+3H^{2}&=&-\kappa^{2}P_{(eff)}
\eea
where
\bea
\label{setens2}
\rho_{(eff)}&=&\frac{1}{f^{'}(R)}
\left[\rho^{(m)}+\frac{Rf^{'}(R)-f(R)}{2}-3H\dot{R}f^{''}(R)\right]\nn \\
P_{(eff)}&=&\frac{1}{f^{'}(R)}
\left[P^{(m)}+f^{'''}(R)(\dot{R})^{2}+2H\dot{R}f^{''}(R)+\right.\nn \\
&+&\left.\ddot{R}f^{''}(R)+\frac{f(R)-Rf^{'}(R)}{2}\right]
\eea
Also we have $\Box :=g^{\mu\nu}\nabla_{\mu}\nabla_{\nu}=-\partial_{tt}-3H\partial_{t}$
and thus using Eqs. (\ref{setens2}) one can easily compute that
\bea
\label{boxf}
\Box f^{'}(R)&=&\left(P^{(m)}-\frac{1}{3}\rho^{(m)}\right)-
f^{'}(R)\left(P_{(eff)}-\frac{1}{3}\rho_{(eff)}\right)+\nn \\
&+&\frac{2}{3}\left[f(R)-Rf^{'}(R)\right]
\eea
From Eqs. (\ref{fieldeq2}) we have
\bea
\label{effcons}
\dot{\rho}_{(eff)}=-3H(\rho_{(eff)}+P_{(eff)})\Longrightarrow
P_{(eff)}=-\frac{\dot{\rho}_{(eff)}}{3H}+\rho_{(eff)}
\eea
Then Eq. (\ref{trace}) becomes
\bea
\label{trace1}
T+f^{'}(R)
\left[\frac{\dot{\rho}_{(eff)}}{H}+2\rho_{(eff)}\right]-Rf^{'}(R)=\kappa^{2}T
\eea
and the first of Eqs. (\ref{fieldeq2}) becomes
\bea
\label{friedman1}
f^{'}(R)
\left[3H^{2}-\frac{\kappa^{2}R}{2}+\frac{\kappa^{2}\rho^{'}_{(eff)}}{2}\frac{a}{a^{'}}\right]=\frac{1}{2}\kappa^{2}(\kappa^{2}-1)T
\eea
while we will be using Eq. (\ref{dust}) also.

%The central assumption of this paper in now that
%$P_{(eff)}:=w_{f}\rho_{(eff)}$ and along with
%$T_{(eff)}=(3w_{f}-1)\rho_{(eff)}$, Eq. (\ref{trace}) becomes
%\bea
%\label{trace1}
%T-f^{'}(R)T_{(eff)}-Rf^{'}(R)=\kappa^{2}T
%\eea
%and the first of Eqs. (\ref{fieldeq2}) becomes
%\bea
%\label{friedman1}
%f^{'}(R)
%\left[3H^{2}+\frac{\kappa^{2}R}{3w_{f}-1}\right]+\kappa^{2}(\kappa^{2}-1)T=0
%\eea
%Also from the stress-energy tensor conservation, along with
%$P^{(m)}:=w\rho^{(m)}$ we have
%\bea
%\label{trace2}
%T=(3w-1)\rho^{(m)}=(3w-1)\rho_{0}
%\left(\frac{a_{0}}{a}\right)^{3(1+w)}
%\eea

We now seek for a {\it bouncing} cosmological model, so that
for $t=0$ we demand that the scalar curvature is finite,
$R(t=0):=R_{0}>0$ and moreover we must have
$\dot{R}(t=0)=\dot{R}(R=R_{0})=0$. This indeed
implies, from $\dot{a}=a^{'}(R)\dot{R}$, that $\dot{a}(t=0)=0$,
while in general $a^{'}(R=R_{0})\neq 0$.
Also $a_{0}:=a(t=0)=a(R=R_{0})>0$ must be finite. Moreover we
seek for a late time {\it accelerating} cosmology. Thus for
$t\longrightarrow\infty$ we must have $\dot{H}_{\infty}=0$ and
thus
$R\longrightarrow R_{\infty}=6(\dot{H}_{\infty}+2H_{\infty}^{2})=12H_{\infty}^{2}>0$.
Then from Eq. (\ref{scalar}) we obtain
\bea
\label{scalarde}
\frac{d}{dR}\left[(\dot{R})^{2}\right]+
2\left[\frac{a^{''}}{a}+\frac{a^{'}}{a}\right](\dot{R})^{2}=
\frac{Ra}{3a^{'}}
\eea
with solution
\bea
\label{solution}
(\dot{R})^{2}=\frac{1}{3(aa^{'})^{2}}
\int_{R_{0}}^{R}Ra^{3}a^{'}(R)dR
\eea
Now we compute from Eq. (\ref{solution}), the term $3H^{2}$ and also
use Eq. (\ref{dust})
into Eq. (\ref{friedman1}). We then take a derivative with respect to $R$,
and also we assume, as stated previously, that
the matter content of the Universe is in the form of pressureless
dust, namely we set $w=0$. After some algebra we finally obtain
\bea
\label{friedman2}
\left(\frac{a^{'}}{a}\right)&=&\frac{\kappa^{2}(\kappa^{2}-1)\rho_{0}}{2Rf^{'}(R)}
\left(\frac{a_{0}}{a}\right)^{3}\left(\frac{a^{'}}{a}\right)-
\frac{\kappa^{2}(\kappa^{2}-1)\rho_{0}}{2[f^{'}(R)]^{2}}
\left(\frac{a_{0}}{a}\right)^{3}\frac{f^{''}(R)}{R}+\nn \\
&+&\frac{\kappa^{2}}{2R}+2\kappa^{2}\left(\frac{a^{'}}{a}\right)
-\frac{5\kappa^{2}\rho^{'}_{(eff)}(R)}{2R}-
\frac{\kappa^{2}}{2}\left(\frac{a}{a^{'}}\right)\frac{\rho^{''}_{(eff)}}{R}+\nn
\\&+&\frac{\kappa^{2}}{2}\frac{\rho^{'}_{(eff)}(R)}{R}\frac{aa^{''}}{(a^{'})^{2}}
\eea
We now normalize the curvature with respect to $R_{0}$, namely
we define the new variable $R_{n}:=R/R_{0},\; \; \; 0<R_{n}\leq 1$.
Also it is assumed from now on that a prime denotes derivative with
respect to the new normalized variable for the curvature, namely
$()^{'}:=d()/dR_{n}$. Also we define
$Y:=(a_{0}/a)^{3}$, $c_{0}:=\frac{\kappa^{2}}{2(2\kappa^{2}-1)}$,
$R_{\rho}:=\kappa^{2}\rho_{0}/2$ and $c_{1}:=\frac{(\kappa^{2}-1)}{(2\kappa^{2}-1)}$.
It is evident that $R_{\rho}$ is a measure (in units of curvature)
of the mass-energy density of the Universe,
at the moment of bouncing. Then we make our unique central assumption
in this paper, that specifies the class of solutions that we seek to study,
in the form of $\rho_{(eff)}:=\rho_{1}Y^{m}$ where $m, \;\rho_{1}$ are
constants to be determined below, Eq. (\ref{friedman2}) becomes
\bea
\label{yder}
Y^{'}=
\frac{3Y\left[c_{0}-\frac{c_{1}R_{\rho}}{f^{'}(R_{n})}\frac{f^{''}}{f^{'}}Y\right]}
{R_{n}\left[1+\frac{c_{1}R_{\rho}}{R_{n}f^{'}(R_{n})}Y\right]}
\eea
In terms of the above definitions Eq. (\ref{solution})
becomes
\bea
\label{solution1}
(\dot{R}_{n})^{2}=\frac{R_{0}Y^{10/3}}{(Y^{'})^{2}}
\int_{R_{n}}^{1}\frac{R_{n}Y^{'}}{Y^{7/3}}dR_{n}:=\frac{R_{0}GY}{(Y^{'})^{2}}\nn
\\
G:=G(R_{n},Y)=\frac{3}{4}Y^{7/3}
\left[\frac{R_{n}}{Y^{4/3}}-1+\int_{R_{n}}^{1}\frac{dR_{n}}{Y^{4/3}}\right]
\eea
while from the first of Eqs. (\ref{fieldeq1}) and using our
assumption $\rho_{(eff)}:=\rho_{1}Y^{m}$ we obtain
\bea
\label{rho1}
\rho_{0}Y+\frac{1}{2}[R_{n}f^{'}(R_{n})-f(R_{n})]+
\frac{Y^{'}}{Y}\frac{(\dot{R}_{n})^{2}}{R_{0}}f^{''}(R_{n})=
\frac{\rho_{1}}{R_{0}}f^{'}(R_{n})Y^{m}
\eea

\section{Reduction of the system and analytical treatment}

In principle we would have the freedom to assume a quite general
form for the $f(R)-$gravity models under
study. For example we would assume that we have the following
class of polynomial $f(R)-$gravity models, of the form
\bea
\label{fR}
f(R)=-2\Lambda+f_{1}R+\frac{1}{2}\frac{f_{2}}{R_{0}}R^{2}+\frac{1}{3}\frac{f_{3}}{R_{0}^{2}}R^{3}
\eea
Here $\Lambda>0$ is the cosmological constant, $f_{1},\; f_{2},\; f_{3}$ are
dimensionless constants and $R_{0}$ is the finite curvature scalar at the
moment of bouncing, which may be taken to be of the order of the inverse
square of the Planck length, namely
\bea
\label{R0}
R_{0}\simeq \frac{1}{L_{pl}^{2}}\simeq 0.385\times 10^{70} m^{-2}
\eea
Also we can make an assumption for the order of magnitude estimate
of
\bea
\label{Rrho}
R_{\rho}=\frac{8\pi}{2}\frac{M_{pl}}{L_{pl}^{3}}
\simeq 4.835\times 10^{70}m^{-2}
\eea
and $\lambda:=R_{\rho}/R_{0}\simeq 12.563$. The dimension of the function
$f(R)$ in Eq. (\ref{fR}) is that of a scalar curvature, namely that of
$R_{0}$, for example, so we define the completely dimensionless function
$F(R_{n}):=f(R_{n})/R_{0}$ and Eq. (\ref{yder}) assumes its completely
dimensionless-normalized form, suitable for numerical treatment, as follows
\bea
\label{yder1}
Y^{'}=
\frac{3Y\left[c_{0}-\frac{c_{1}\lambda}{F^{'}(R_{n})}\frac{F^{''}}{F^{'}}Y\right]}
{R_{n}\left[1+\frac{c_{1}\lambda}{R_{n}F^{'}(R_{n})}Y\right]}
\eea
where of course from above we obtain $\lambda\simeq 12.563$,
$c_{0}=0.255$ and $c_{1}=0.49$, due to the fact that we have chosen units
where $\kappa^{2}=8\pi$. Also we define the free dimensionless parameter
$\mu:=\rho_{1}/R_{0}$. Then Eq. (\ref{rho1}) assumes its completely
dimensionless-normalized form as
\bea
\label{rho11}
\lambda Y+2\pi [R_{n}F^{'}(R_{n})-F(R_{n})]+\frac{G}{Y^{'}}F^{''}(R_{n})=
\mu F^{'}(R_{n})Y^{m}
\eea
Finally using Eqs. (\ref{yder1}), (\ref{solution1}) into Eq. (\ref{rho11}) we obtain
after some algebra, the completely dimensionless-normalized form for the
evolution of the $f(R)-$gravity curvature function as
\bea
\label{fdder}
F^{''}&=&\frac{G_{1}}{G_{2}}\nn \\
G_{2}&:=&G_{2}(R_{n},Y,F,F^{'})=G R_{n}
\left[1+\frac{c_{1}\lambda}{R_{n}F^{'}}Y\right]+\nn \\&+&
3Y^{2}\frac{c_{1}\lambda}{(F^{'})^{2}}
\left[\mu F^{'}Y^{m}-\lambda Y-2\pi [R_{n}F^{'}-F]\right]\nn
\\
G_{1}&:=&G_{1}(R_{n},Y,F,F^{'})=3c_{0}Y
\left[\mu F^{'}Y^{m}-\lambda Y-2\pi [R_{n}F^{'}-F]\right]
\eea
Thus we can numerically integrate Eqs. (\ref{yder1}), (\ref{fdder}),
using also the last of Eqs. (\ref{solution1}).

Now we study the Hubble parameter of these models.
Regarding the Hubble parameter in our epoch,
in the notation used here, it is given by [\citen{kinn}]
\bea
\label{Hubble}
H_{\infty}=72\pm 8 Km/sec/Mpc
\eea
Assuming that the Universe in the present moment is accelerating
%in a quasi-De-Sitter fashion
we obtain
$R_{\infty}=6(\dot{H}_{\infty}+2H_{\infty}^{2})=12H_{\infty}^{2}=0.726\times 10^{-39}m^{-2}$.
When expressed in normalized-dimensionless units, with respect to the
curvature constant $R_{0}$ of Eq. (\ref{R0}), it assumes the form
\bea
\label{Rinf}
{\cal R}_{\infty}:=\frac{R_{\infty}}{R_{0}}\simeq 1.885\times 10^{-109}
\eea
A careful calculation, using Eqs. (\ref{scalarde}), (\ref{solution1}) yields
\bea
\label{dHH2}
A:=1+\frac{\dot{H}}{H^{2}}=1+
\frac{\left[1-\int_{R_{n}}^{1}\frac{dR_{n}}{Y^{4/3}}\right]}
{2\left[\frac{R_{n}}{Y^{4/3}}-1+\int_{R_{n}}^{1}\frac{dR_{n}}{Y^{4/3}}\right]}
\eea
A sufficient condition for late time acceleration is that as
$R_{n}\longrightarrow 0$ we have $A>0$.

\begin{figure}[h!]
\centerline{\includegraphics[width=10.8cm]{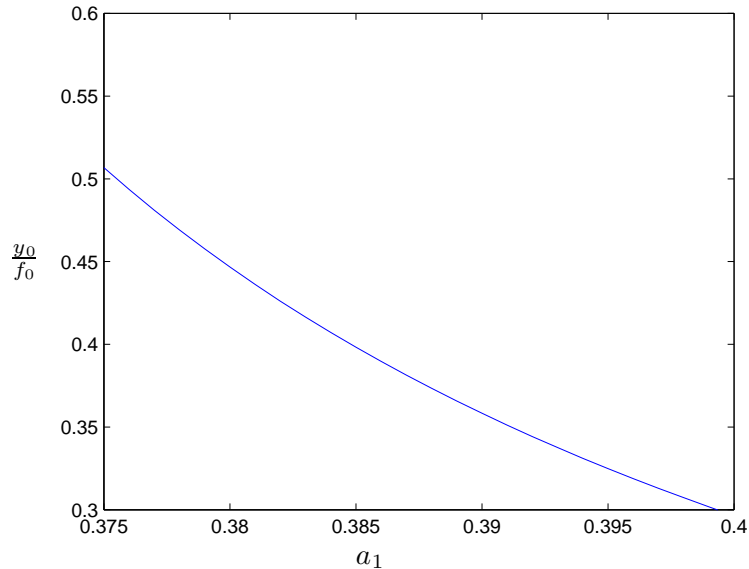}}
\caption{The function of Eq. (\ref{r}), plotted against the permitted range
of parameter values of $a_{1}$, given by Eq. (\ref{a1}).}
\label{fig1}
\end{figure}

\begin{figure}[h!]
\centerline{\includegraphics[width=10.8cm]{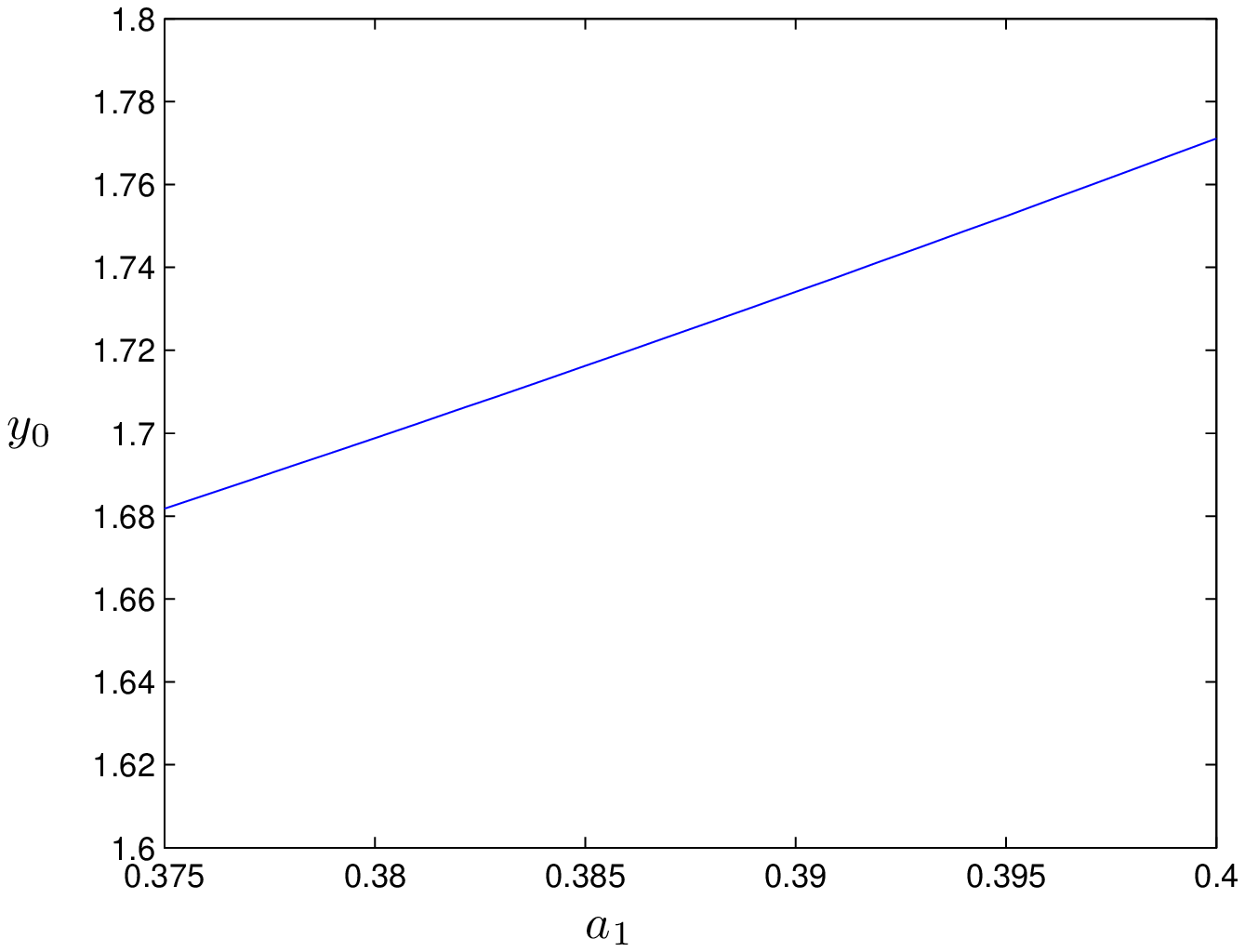}}
\caption{The function of Eq. (\ref{y0}), plotted against the permitted range
of parameter values of $a_{1}$, given by Eq. (\ref{a1}).}
\label{fig2}
\end{figure}

\begin{figure}[h!]
\centerline{\includegraphics[width=10.8cm]{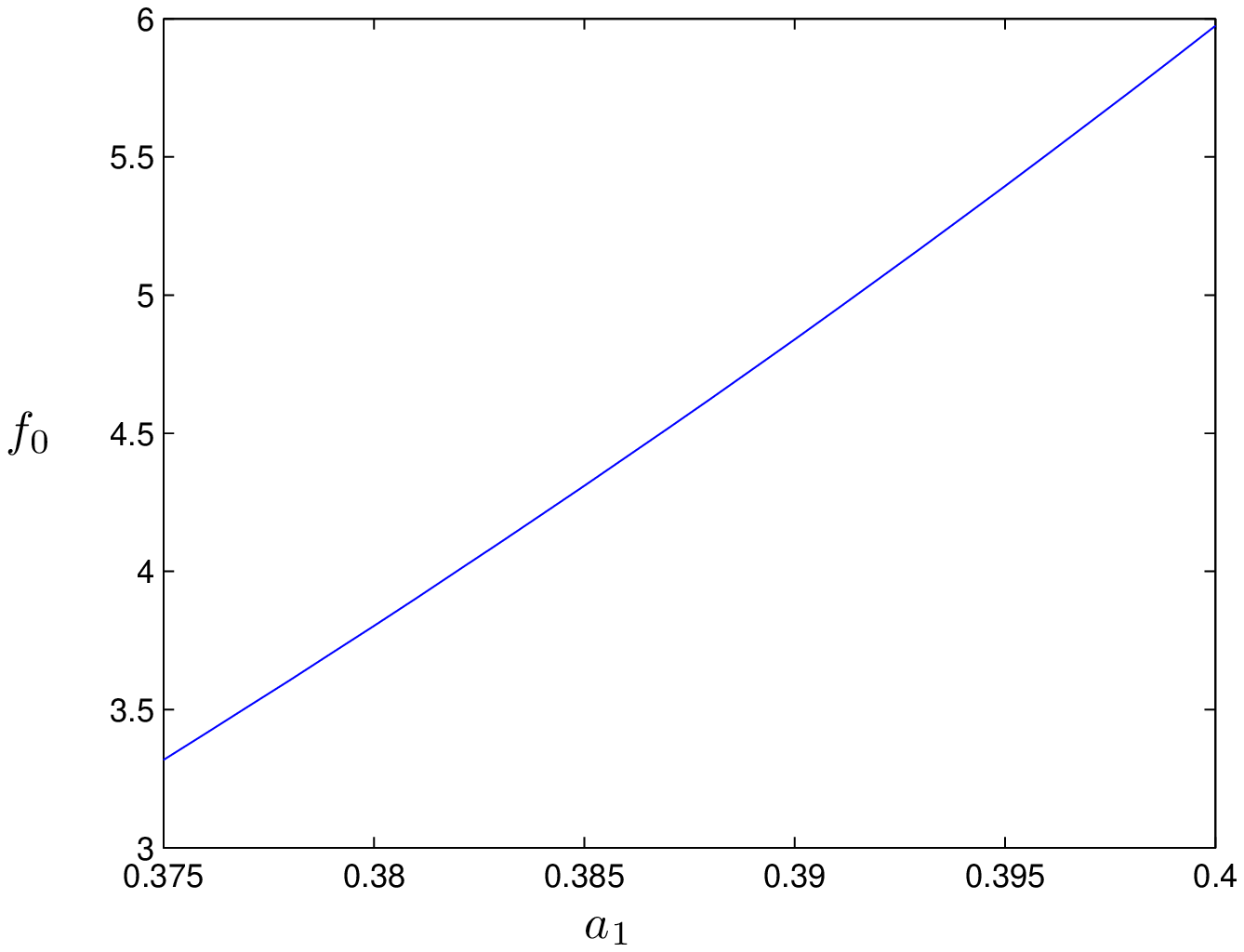}}
\caption{The constant $f_{0}$ of Eq. (\ref{YF}), plotted against the permitted range
of parameter values of $a_{1}$, given by Eq. (\ref{a1}).}
\label{fig3}
\end{figure}

\begin{figure}[h!]
\centerline{\includegraphics[width=10.8cm]{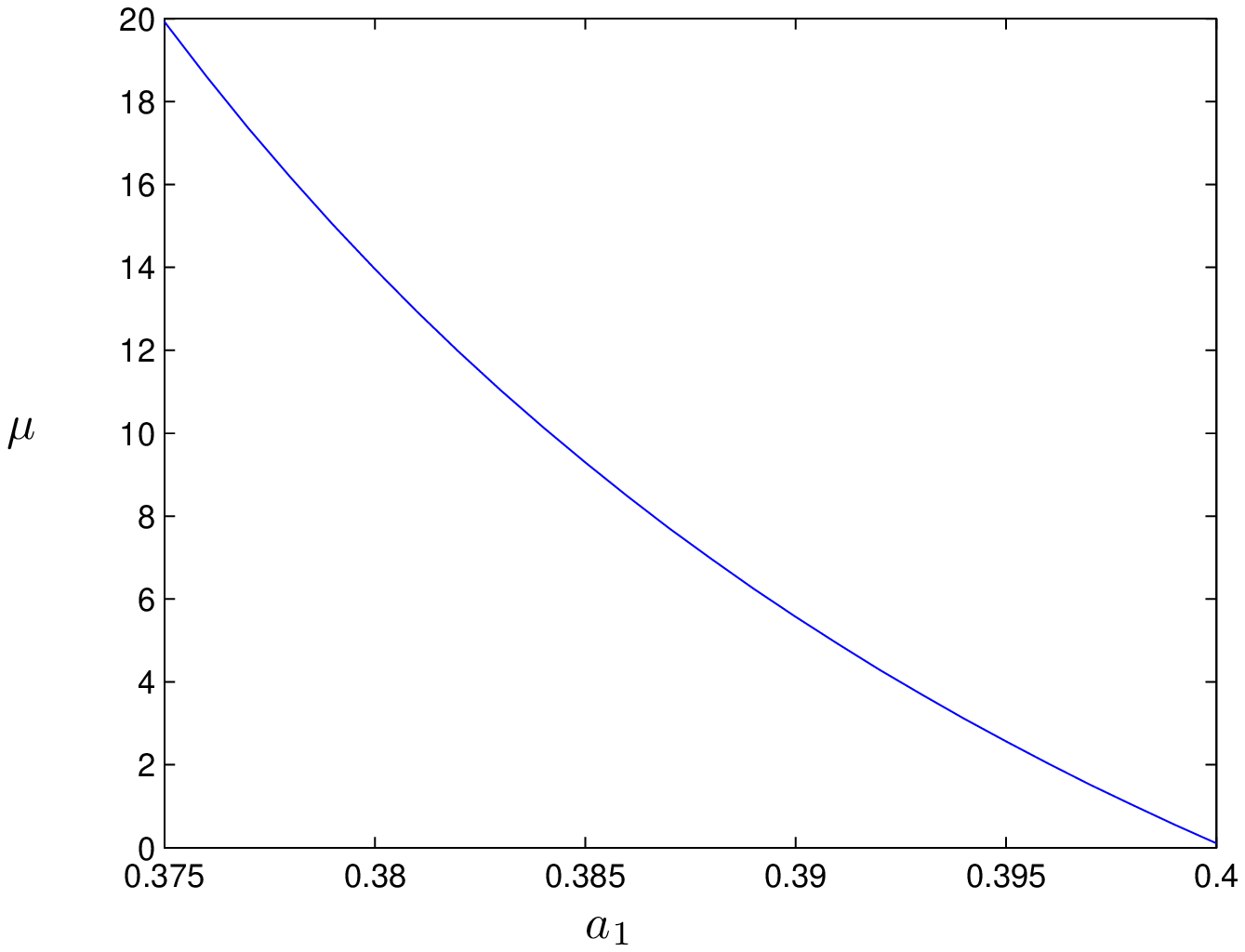}}
\caption{The constant $\mu>0$ of Eq. (\ref{mu}), plotted against the permitted range
of parameter values of $a_{1}$, given by Eq. (\ref{a1}).}
\label{fig4}
\end{figure}

We now present a class on exact, analytical, cosmological models
which are bouncing at the origin of time $t=0$ (corresponding to $R_{n}=1$),
and also are late-time accelerating. We assume the following form
of the functions $Y,\; F$
\bea
\label{YF}
Y&:=&y_{0}R_{n}^{a_{1}}\nn \\
F&:=&f_{0}R_{n}^{a_{1}}
\eea
where $y_{0},\; f_{0},\; a_{1}$ are positive constants. Then from
Eq. (\ref{yder1}) we obtain
\bea
\label{r}
r:=\frac{y_{0}}{f_{0}}=\frac{3c_{0}-a_{1}}{c_{1}\lambda (3a_{1}-1)}
\eea
This is plotted in Fig. \ref{fig1}, for the allowed parameter region
of the constant $a_{1}$, to be determined below. We then substitute
the first of Eqs. (\ref{YF}) into Eq. (\ref{dHH2}) and
after a careful evaluation we find that in order to have late-time accelerating
cosmologies, namely to have $A>0$ as $R_{n}\longrightarrow 0$ it is necessary that
$\alpha_{1}>3/8$. Then we substitute the first of Eqs. (\ref{YF}) into the
definition of the function $G$, of Eq. (\ref{solution1}) and we choose
to set
\bea
\label{y0}
y_{0}^{4/3}:=\frac{1}{\left(1-\frac{4a_{1}}{3}\right)}>0
\eea
This is plotted in Fig. \ref{fig2}, for the allowed parameter region
of the constant $a_{1}$, to be determined below.
From Eqs. (\ref{dHH2}), (\ref{y0}) we thus find that the range of
permitted values of the constant $a_{1}$ is
\bea
\label{a1}
\frac{3}{8}<a_{1}<\frac{3}{4}
\eea
The constant $f_{0}$ is readily determined by the combination of
Eqs. (\ref{r}), (\ref{y0}) and it is plotted in Fig. \ref{fig3}.
All the previous results are now substituted into Eq. (\ref{fdder}).
Also we assume that the constant $m=1/a_{1}$. We then find after
some careful algebra that
\bea
\label{mu}
& &3\mu \left[rc_{1}\lambda(a_{1}-1)-c_{0}a_{1}\right]y_{0}^{m}=
a_{1}(a_{1}-1)(a_{1}+c_{1}\lambda r)+\nn \\
&+&
\left[\frac{3rc_{1}\lambda (a_{1}-1)}{a_{1}}-3c_{0}\right]
\left[\lambda r+2\pi (a_{1}-1)\right]
\eea
This determines the positive constant $\mu$ and it
is plotted in Fig. \ref{fig4}. Thus Eqs. (\ref{YF}) constitute an exact
analytic cosmological model for the range of the parameter $a_{1}$, given
by Eq. (\ref{a1}). These are bouncing at the origin of time ($t=0$), as
it will also be explained in Section 6, and are late-time accelerating.
This model shows that one can obtain relatively easy, simple cosmological
models with the aforementioned properties,
in the frame of the $f(R)-$gravity theories. Using Eq. (\ref{Rinf}) into
the first of Eqs. (\ref{YF}) we obtain
\bea
\label{efolds}
N:=ln\left(\frac{a_{\infty}}{a_{0}}\right)\simeq 251 a_{1}
\eea
for the number of e-foldings from the initial moment of the Universe
up to the present day. The minimum value thus is of the order of
$N_{min}\simeq 94$, well within acceptable values required by
inflation theory [\citen{kinn}]  .

%Then from Eq. (\ref{solution2}) we obtain for the value of the scale factor
%in the present moment $a_{\infty}$, relative to its minimum value $a_{0}$, that
%\bea
%\label{scalef}
%\frac{a_{\infty}}{a_{0}}\simeq 7.617\times 10^{36}\Longrightarrow
%N:=ln\left(\frac{a_{\infty}}{a_{0}}\right)\simeq 85
%\eea
%where $N$ is the number of e-foldings.

\section{Numerical treatment}

\begin{figure}[h!]
\centerline{\includegraphics[width=10.8cm]{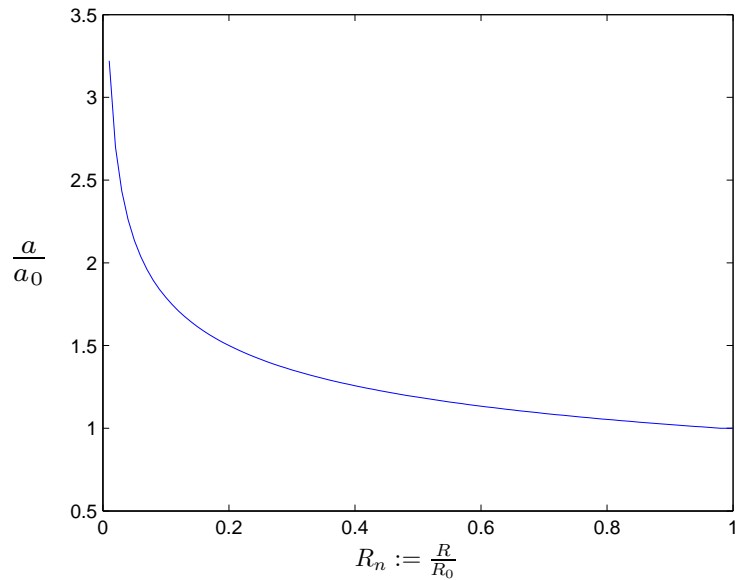}}
\caption{The scale factor as a function of the normalized scalar curvature
for the numerically determined cosmological model of Eq. (\ref{fit1}).}
\label{fig5}
\end{figure}

\begin{figure}[h!]
\centerline{\includegraphics[width=10.8cm]{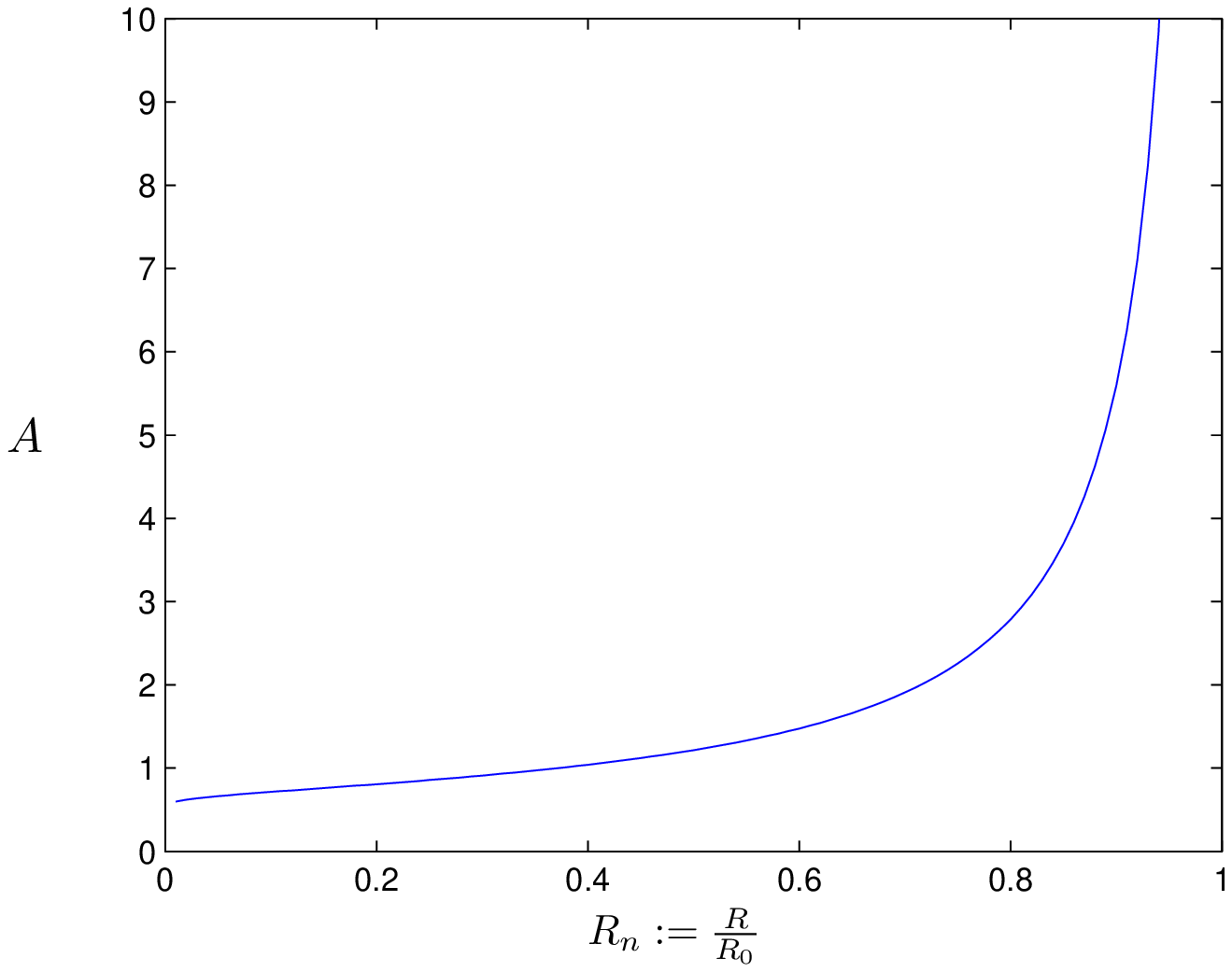}}
\caption{The function $A$, of Eq. (\ref{dHH2})
as a function of the normalized scalar curvature
for the numerically determined cosmological model of Eq. (\ref{fit1}).}
\label{fig6}
\end{figure}

\begin{figure}[h!]
\centerline{\includegraphics[width=10.8cm]{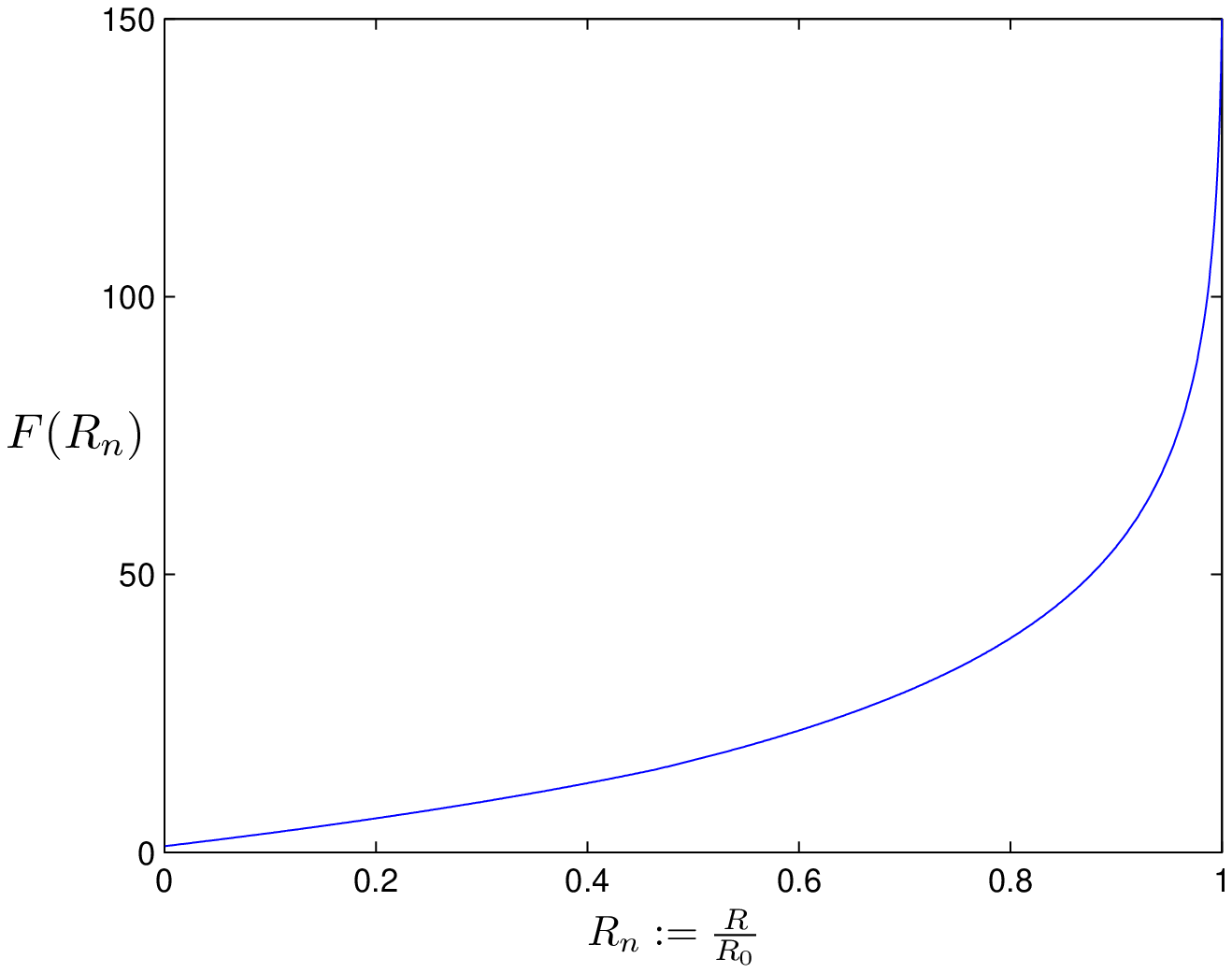}}
\caption{The function $F(R_{n})$, of Eq. (\ref{fit2})
as a function of the normalized scalar curvature
for the numerically determined cosmological model of Eq. (\ref{fit1}).}
\label{fig7}
\end{figure}

As an explicit example of the possibility of numerical treatment
of the set of Eqs. (\ref{yder1}), (\ref{fdder}) we give the following
result. We have integrated numerically Eqs. (\ref{yder1}), (\ref{fdder}),
with starting point $R_{n}=1$, backwards, towards $R_{n}\longrightarrow 0$,
where presumably the present epoch occurs. We used fourth order Runge-Kutta
method with step size as small as $h=10^{-3}$. The initial conditions
used were $Y(1)=1$, $F(1)=150$, $F^{'}(1)=10^{4}$. The values of the freely
specified constants were assumed to be $\mu=1.1$, $m=1$. The result is
shown in Figs. \ref{fig5}, \ref{fig6}, \ref{fig7}. Using fitting of the
numerical data we have found that for Fig. \ref{fig5}
\bea
\label{fit1}
Y\sim R_{n}^{d},\; \; \; d:=2.191
\eea
while for Fig. \ref{fig7} we found
\bea
\label{fit2}
F(R_{n})\sim F_{1} R_{n}+F_{2}R_{n}^{2}+F_{3}R_{n}^{3},\nn \\
F_{1}:=109.6,\; \; F_{2}:=-276,\; \;F_{3}:=262.4
\eea
Finally from Fig. \ref{fig6} we see that it is a late-time accelerating
cosmology and using Eq. (\ref{efolds}) the number of e-foldings is
also well within acceptable values required by inflation theories.

%\newpage
%Finally we assume that
%\bea
%\label{wf}
%1+\frac{4\kappa^{2}}{3w_{f}-1}=0.
%\eea
%Then Eq. (\ref{friedman1}) becomes
%\bea
%\label{friedman2}
%\frac{dY}{dR}+\frac{3f^{''}(R)}{f^{'}(R)}Y=\frac{f^{'}(R)}{R_{\rho}}
%\eea
%where we have defined $Y:=(a_{0}/a)^{3}$ and $R_{\rho}:=4\kappa^{2}(\kappa^{2}-1)\rho_{0}/3$.
%It is evident that $R_{\rho}$ is a measure (in units of curvature)
%of the mass-energy density of the Universe,
%at the moment of bouncing. The solution to Eq. (\ref{friedman2}) is given by
%\bea
%\label{solution1}
%Y=\frac{Y_{0}+\frac{1}{R_{\rho}}\int[f^{'}(R)]^{4}dR}{[f^{'}(R)]^{3}}
%\eea
%where the integration constant $Y_{0}$ is set to zero, for
%reasons that will become clear in what follows.

\section{Smoothness of solutions}

%We now have the freedom to assume a quite general form for the $f(R)-$gravity models under
%study. We thus assume that we have the following class of polynomial $f(R)-$gravity
%models, namely
%\bea
%\label{fR}
%f(R)=-2\Lambda+f_{1}R+\frac{1}{2}\frac{f_{2}}{R_{0}}R^{2}+\frac{1}{3}\frac{f_{3}}{R_{0}^{2}}R^{3}
%\eea
%Here $\Lambda>0$ is the cosmological constant, $f_{1},\; f_{2},\; f_{3}$ are
%dimensionless constants and $R_{0}$ is the finite curvature scalar at the
%moment of bouncing, which may be taken to be of the order of the inverse
%square of the Planck length, namely
%\bea
%\label{R0}
%R_{0}\simeq \frac{1}{L_{pl}^{2}}\simeq 0.385\times 10^{70} m^{-2}
%\eea
%Also we can make an assumption for the order of magnitude estimate
%of
%\bea
%\label{Rrho}
%R_{\rho}=\frac{4\times8\pi\times(8\pi-1)}{3}\frac{M_{pl}}{L_{pl}^{3}}
%\simeq 311.16\times 10^{70}m^{-2}
%\eea

%scale factor of Eq. (\ref{metric}), as it follows from Eq. (\ref{solution1}),
%is finally given by
%\bea
%\label{solution2}
%Z:=\frac{a}{a_{0}}=\frac{f^{'}(R_{n})}{[\lambda I(R_{n};f_{1},f_{2},f_{3})]^{1/3}}
%\eea
%where $\lambda:=R_{0}/R_{\rho}\simeq 0.0012$
%and the function $I$ is given in the Appendix.

\begin{figure}[h!]
\centerline{\includegraphics[width=10.8cm]{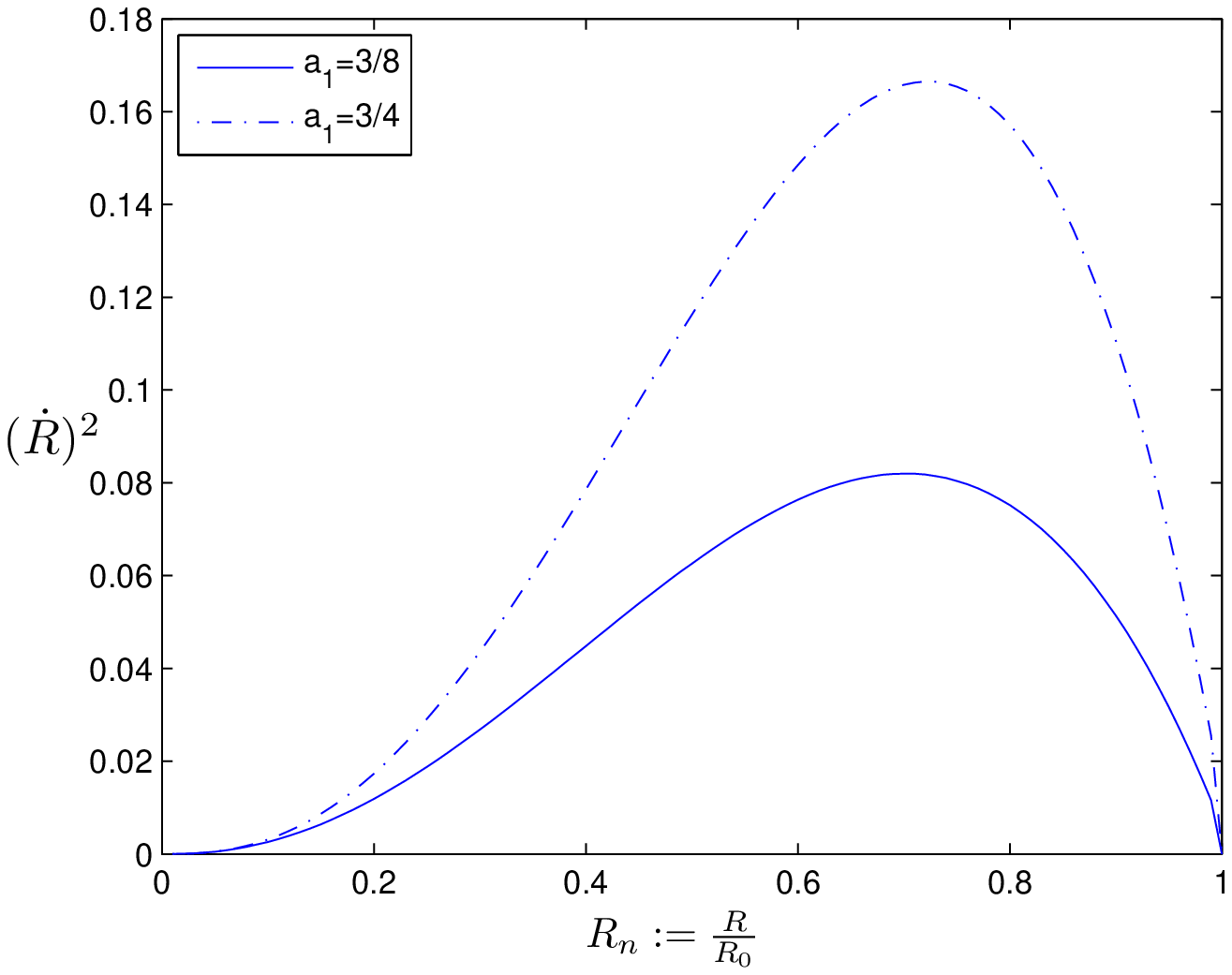}}
\caption{The function of Eq. (\ref{solution}), plotted against the
normalized scalar curvature, for permitted range
of parameter values of $a_{1}$, given by Eq. (\ref{a1}).}
\label{fig8}
\end{figure}

%As a concrete example for $f_{1}=1$, $f_{2}=0.81$, $f_{3}=-1.0023$ we obtain
%the scale factor of Fig. \ref{fig1}. The time derivative of the scalar curvature
%of Eq. (\ref{solution}) is shown in Fig. \ref{fig2}.
Now a crucial point
is to elucidated here. In the construction of the cosmological models
appearing in this paper,
we consider Eq. (\ref{solution}) with the positive sign for the root, namely
\bea
\label{solution3}
(\dot{R})=\frac{1}{\sqrt{3}(aa^{'})}
\left[\int_{R_{0}}^{R}Ra^{3}a^{'}(R)dR\right]^{1/2}
\eea
This is because we assume that the scale factor decreases with increasing
scalar curvature (for the case of the expanding phase of the Universe,
namely for the case $0\leq t<+\infty$), and so $a^{'}(R)<0$, implying that
we also have $\dot{R}<0$. Now formally we would solve Eq. (\ref{solution3}),
with the initial condition $R(0)=R_{0}>0$ and obtain a solution for the
scalar curvature of the form $R(t)=F(t),\; \; (0\leq t<+\infty)$.
What we want to do here is then to match this
solution with its mirror-symmetric, with respect to the origin
$t=0$, solution $R(t)=G(t)=F(-t),\; \; (-\infty<t\leq 0)$
resulting from Eq. (\ref{solution}), when taken with the negative sign.
We will show that
this matching is {\it smooth}, i.e. the scalar curvature
$R(t),\; \; (-\infty<t<+\infty)$ is of the $C^{(\infty)}-$differentiability class.
Indeed we can easily demonstrate the equality of the two-side,
even order, derivatives
\bea
\label{evend}
\left.\frac{d^{2k}R}{dt^{2k}}\right|_{0^{+}}=
\left.\frac{d^{2k}F}{dt^{2k}}\right|_{0^{+}}=
\left.\frac{d^{2k}G}{dt^{2k}}\right|_{0^{-}}=
\left.\frac{d^{2k}R}{dt^{2k}}\right|_{0^{-}}
\eea
Also from Eq. (\ref{solution3}) we can relatively easy show that
all the two-side odd order derivatives of the scalar curvature $R(t)$ vanish,
at $t=0$, or equivalently at $R=R_{0}$, namely
\bea
\label{oddd}
\dot{R}(0)=\dddot{R}(0)=...=R^{(2k+1)}(0)=0
\eea
and thus one proves that the constructed function of the scalar curvature
$R(t),\; \;(-\infty<t<+\infty)$ is everywhere smooth.
From Eq. (\ref{yder1}) this implies that the scale factor $a(R(t))$ is
also smooth. As an additional argument to support the claim of this section,
in Fig. \ref{fig8} it is plotted the function of Eq. (\ref{solution}), for the
case of the cosmological models of Eqs. (\ref{YF}) and for the permitted range
of values for the parameter $a_{1}$, appearing in Eq. (\ref{a1}).

%\newpage

\section{Discussion}

We have considered the field equations for a flat FRW cosmological model
in the frame of $f(R)$ gravity. We have reduced them into a completely
normalized-dimensionless system of O.D.Es for the scale factor of the
Universe and the function $f(R)$. Then through an ansatz we have produced
a set of simple analytical cosmological models which are bouncing at the
origin and late-time accelerating also. This occurs for certain values
of the relevant parameters involved. Also a numerical treatment of this
system results in similar cosmological models. It would be interesting
to try to generalize and implement this construction for more general
cosmological models and extended theories of gravity. Work along these
lines is in progress.

\section*{Acknowledgments}

The author would like to acknowledge useful discussions with
Dr. K. Kleidis and other colleagues in the Technological Education
Institute (T.E.I.) of Central Macedonia, Greece.

\end{document}